\documentclass[draftcls, onecolumn, 12pt]{IEEEtran}
\usepackage{latexsym,bm}
\usepackage[tbtags]{amsmath}
\usepackage[dvips]{graphicx,color}
\usepackage{subfigure}
\usepackage{amssymb}
\usepackage{cite}
\usepackage{stfloats}

\usepackage{times}
\usepackage{caption}
\usepackage{amsmath}
\usepackage{amssymb}
\usepackage{amscd}
\usepackage{cite}
\usepackage{amsfonts}
\usepackage{graphicx}
\usepackage{psfrag}
\usepackage{epsfig}
\usepackage{algorithm}
\usepackage{algorithmic}
\usepackage{flushend}
\ifCLASSINFOpdf
\else
\fi
\begin{document}
%
\title{Transceiver Design to  Maximize Sum Secrecy Rate in Full Duplex SWIPT Systems}
\author{ Ying Wang, ~\IEEEmembership{Member,~IEEE,}
         Ruijin Sun,
         and Xinshui Wang
\thanks{This paragraph of the first footnote will contain the  date on which you submitted your paper for review.  This work was supported by  National 863 Project under grant 2014AA01A701 and National Nature Science Foundation of China(61431003, 61421061).  }
\thanks{Y. Wang, R. Sun and X. Wang are with the State Key Laboratory of Networking and Switching Technology, Beijing University of Posts and Telecommunications, Beijing 100876, P.R. China
(email:  wangying@bupt.edu.cn, sunruijin1992@gmail.com, wxinshui@126.com). }
\thanks{ }}
\IEEEaftertitletext{\vspace{-1\baselineskip}}

\maketitle

\begin{abstract}

 This letter considers secrecy simultaneous wireless information and power transfer (SWIPT) in full duplex systems. In such a system,  full duplex capable base station (FD-BS) is designed to transmit data to one downlink user  and concurrently receive data from one uplink user, while one idle user  harvests the radio-frequency (RF) signals energy to extend its lifetime. Moreover,  to prevent eavesdropping,  artificial noise (AN) is exploited by FD-BS to degrade the channel of  the idle user, as well as to provide energy supply to the idle user.  To maximize the sum of downlink secrecy rate and uplink secrecy rate, we jointly optimize the information covariance matrix, AN covariance matrix and receiver vector, under the constraints of the sum transmission power of FD-BS and the minimum harvested energy of the idle user. Since the problem is non-convex, the log-exponential reformulation and sequential parametric convex approximation (SPCA) method are used.   Extensive simulation results are provided and demonstrate that our proposed full duplex scheme extremely outperforms the half duplex scheme.
\end{abstract}


%
\IEEEpeerreviewmaketitle

\section{Introduction}

\IEEEPARstart{F}{ull-duplex} (FD), potentially doubling the spectral efficiency, has gained considerable attention. Nonetheless, simultaneous information transmission and reception make FD transceivers suffer from the self-interference
(SI) from transmit antennas to receive antennas. Fortunately,
in recent years, many breakthroughs in hardware design for SI
cancellation (SIC) techniques \cite{bharadia2013full} have effectively suppressed
the SI to the background noise level and thus made FD
communications more practicable. Since then, several studies
regarding FD technology have been conducted, including the SIC schemes \cite{Ahmed2013Self}, new designed communication protocols \cite{zeng2014full, zheng2013improving} and system performance optimization \cite{nguyen2014spectral,zheng2013full,zhu2014joint}.

On the other hand, simultaneous wireless information and power transfer (SWIPT) has emerged  as  an effective solution for saving the energy.   A majority of researches considered the downlink  broadcast SWIPT system consisting of a base station (BS) that
broadcasts signals to a set of users, which are either scheduled
as information decoding receivers (IRs) or energy harvesting
receivers (ERs).   To prevent eavesdropping, artificial noise
(AN) was exploited at the BS to degrade the channel of ERs, as well
as to provide energy supply to ERs \cite{liu2013secrecy, ng2013resource, ng2013multi-objective}.  However, the works above focused on half-duplex (HD) systems which would give rise to a significant loss in spectral efficiency. Moreover, the uplink security cannot be guaranteed  when single antenna uplink users lack the required spatial degrees of freedom to ensure secure communication. Multiple-antenna  full duplex capable base station (FD-BS) is a promising solution. With simultaneous transmission and reception, not only the downlink but also the
uplink wiretap channel can be concurrently degraded by the AN transmitted by FD-BS. Another advantage is that ERs in FD systems can harvest the energy from both the downlink and uplink signals in each time slot.

Motivated by the discussion above, in this letter, we study the secure transmission in full duplex
SWIPT systems. Different from the downlink secrecy rate maximization subject to the uplink secrecy rate constraint in full
duplex systems \cite{zhu2014joint}, we maximize the sum of downlink secrecy
rate and uplink secrecy rate by jointly optimizing the information covariance matrix, AN covariance matrix and receiver
vector. Moreover, the impact of SI and co-channel interference (CCI) is considered. Since the
problem is non-convex, the log-exponential reformulation and
sequential parametric convex approximation (SPCA) method
are used.

\section{System Model and Problem Formulation}

We consider a full duplex wireless communications system for SWIPT as illustrated in Fig. 1. It is assumed that there is one FD-BS, one uplink user ($U_U$), one downlink user ($U_D$) and one idle user ($U_I$) with the capability of RF energy harvesting. The FD-BS communicates with $U_U$  in the uplink channel and $U_D$  in the downlink channel at the same time over the same frequency band. Meanwhile, the idle user  harvests the RF energy broadcasted through the communication process, including the energy emitted by the FD-BS and the uplink user. Suppose that uplink user, downlink user and  idle user are all equipped with a single antenna, while the FD-BS employs $N=N_T+N_R$ antennas, of which $N_T$ transmit antennas are used for transmitting signal in the downlink channel and $N_R$ receive antennas are designed for receiving signal in the uplink channel. We assume that all channels are frequency flat slow fading and  the channel state information (CSI) is known at the FD-BS. It is worth noting that, for the purpose of harvesting more energy, the idle user would like to feedback its CSI to the FD-BS, which also contributes to fight against the eavesdropping.

\begin{figure}
\centering
\includegraphics[width=8.0cm, height=6.5cm]{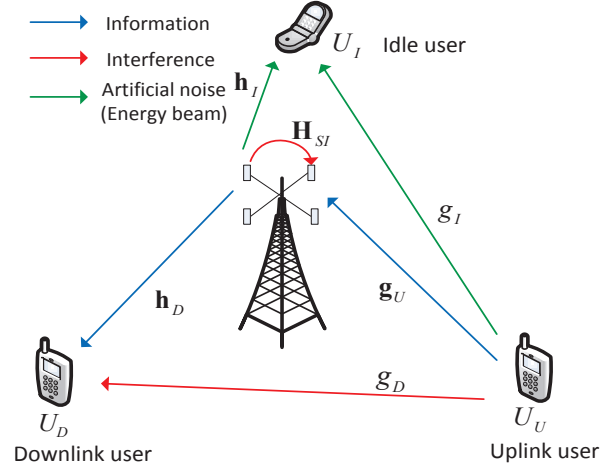}
\caption{Secrecy SWIPT in full duplex system with SI.}
\label{systemmodel}
\end{figure}
To avoid the information leakage, artificial noise is utilized at FD-BS to improve the security from physical layer. Denote the transmit signal sent by FD-BS as
\begin{equation}
{{\mathbf{x}}_D} = {{\mathbf{s}}_D} + {\mathbf{v}},
\end{equation}
where ${{\mathbf{s}}_D} \in {\mathbb{C}^{{N_T} \times 1}}$  is the transmitted data vector intended for $U_D$ which is a complex Gaussian random vector with zero mean and covariance matrix ${\mathbf{S}} \succeq \mathbf{0}$, i.e., ${{\mathbf{s}}_D}\sim {\cal CN}(\mathbf{0},{\mathbf{S}})$.  ${\mathbf{v}} \in {\mathbb{C}^{{N_T} \times 1}}$ is the artificial noise vector generated by the FD-BS to combat the  curious or even adversarial idle user. Similarly, ${\mathbf{v}}\sim {\cal CN}(\mathbf{0},{\mathbf{V}})$, ${\mathbf{V}} \succeq \mathbf{0}$ . Since the idle user harvests energy from the FD-BS, the artificial noise vector also plays the role of energy vector.

Suppose  ${s_U}\sim {\cal CN}(0,1)$ is the data symbol transmitted by uplink user and denote $P_U$ as its corresponding transmission power. Then, the information transmitted by uplink user is given as ${x_U} = \sqrt {{P_U}} {s_U}$.

 The signal received by downlink user $U_D$  and idle user $U_I$  are respectively given by
\begin{equation}
{y_D} = {\mathbf{h}}_D^H{{\mathbf{s}}_D} + {\mathbf{h}}_D^H{\mathbf{v}} + {g_D}{x_U}+ {z_D}
\end{equation}
\begin{equation}
\!\!\!\!\!\!\!\!\!\!\!\!\!\!\text {and}~~
{y_I} = {\mathbf{h}}_I^H{{\mathbf{s}}_D} + {\mathbf{h}}_I^H{\mathbf{v}} + {g_I}{x_U} + {z_I},
\end{equation}
where ${{\mathbf{h}}_D} \in {\mathbb{C}^{{N_T} \times 1}}$ and ${{\mathbf{h}}_I} \in {\mathbb{C}^{{N_T} \times 1}}$ denote the channel vector from FD-BS to user $U_D$  and $U_I$, respectively.  $g_I$ and $g_D$ represent the complex channel coefficient from $U_U$ to $U_I$ and $U_D$, respectively. ${z_D},{z_I}\sim {\cal CN}(0,\sigma _Z^2)$  are the corresponding background noise at $U_D$  and $U_I$, respectively.

After the receiver vector ${{\mathbf{w}}_R}$,  the received signal  at FD-BS is given as
\begin{equation}
{y_U} = {\mathbf{w}}_R^H{{\mathbf{g}}_U}\sqrt {{P_U}} {s_U} +  {{\mathbf{w}}_R^H{{\mathbf{H}}_{SI}}({{\mathbf{s}}_D} + {\mathbf{v}})} + {\mathbf{w}}_R^H{{\mathbf{z}}_U},
\end{equation}
where ${{\mathbf{g}}_U} \in {\mathbb{C}^{{N_R} \times 1}}$ is the complex channel vector from the FD-BS to uplink user $U_U$ and ${\mathbf{z}_U}\sim {\cal CN}(\mathbf{0},\sigma _Z^2{{\mathbf{I}}_{{N_R}}})$ is the noise vector received at the FD-BS.    ${{\mathbf{H}}_{SI}} \in {\mathbb{C}^{{N_R} \times {N_T}}}$ is the residual SI channel from transmit antennas to the receive antennas at FD-BS.

From (2) and (4), the received signal to interference plus noise ratio (SINR) at downlink user $U_D$ and FD-BS can be respectively expressed as
\begin{equation}
{\gamma _D} = \frac{{{\mathbf{h}}_D^H{\mathbf{S}}{{\mathbf{h}}_D}}}
{{{\mathbf{h}}_D^H{\mathbf{V}}{{\mathbf{h}}_D} +{P_U}{{\left| {{g_D}} \right|}^2} + \sigma _Z^2}}
\end{equation}
\begin{equation}
\text{and}~~ {\gamma _U} =
\frac{{{P_U}{{\left| {{\mathbf{w}}_R^H{{\mathbf{g}}_U}} \right|}^2}}}
{{{\mathbf{w}}_R^H{{\mathbf{H}}_{SI}}{\mathbf{(S+V)H}}_{SI}^H{{\mathbf{w}}_R} + \sigma _z^2\left\| {{\mathbf{w}}_R^{}} \right\|_2^2}}.
\end{equation}

 The idle user is assumed to  process the downlink and uplink signals independently \cite{zhu2014joint}. Thereby, from (3), the corresponding SINR of the downlink and uplink signals at idle user are respectively given as
\begin{equation}
{\gamma _I^D} = \frac{{{\mathbf{h}}_I^H{\mathbf{S}}{{\mathbf{h}}_I}}}
{{{\mathbf{h}}_I^H{\mathbf{V}}{{\mathbf{h}}_I} + {P_U}{{\left| {{g_I}} \right|}^2} + \sigma _Z^2}}
\end{equation}
\begin{equation}
\text{and}~~
{\gamma _I^U} = \frac{{P_U}{{\left| {{g_I}} \right|}^2}}
{{{\mathbf{h}}_I^H{\mathbf{V}}{{\mathbf{h}}_I}  +{\mathbf{h}}_I^H{\mathbf{S}}{{\mathbf{h}}_I}+ \sigma _Z^2}}.
\end{equation}
Different from the HD-BS, both downlink and uplink security can be concurrently guaranteed by the AN sent by FD-BS.

 The achievable secrecy rates of downlink and uplink channel  can be respectively expressed as
\begin{align}
  R_D^{\sec }({\mathbf{S}},{\mathbf{V}})&= \Big[{\log _2}(1 + {\gamma _D}) - {\log _2}(1 + {\gamma _I^D})\Big]^+~~\text{and}
\end{align}
\begin{equation}
{R_U^{sec}}({\mathbf{S}},{\mathbf{V}},{{\mathbf{w}}_R}) = \Big[{\log _2}\left( {1 + {\gamma _U}} \right)-{\log _2}\left( {1 + {\gamma _I^U}} \right)\Big]^+,
\end{equation}
where $[x]^+ \triangleq \max\{0,x\}$.

On the other hand, the harvested energy at  $U_I$  is given by
\begin{equation}
E = \zeta \left( {{\mathbf{h}}_I^H{\mathbf{S}}{{\mathbf{h}}_I} + {\mathbf{h}}_I^H{\mathbf{V}}{{\mathbf{h}}_I} + {P_U}{{\left| {{g_I}} \right|}^2}} \right),
\end{equation}
where $0 < \zeta  \leqslant 1$  is a constant,  denoting the RF energy conversion efficiency of the idle user.



 In this letter, we focus on the joint design of transceiver information covariance matrix, AN covariance matrix and receiver vector to maximize the total secure downlink and uplink transmission rate under the sum transmission power constraint at FD-BS and the harvested energy constraint at idle user. Specifically, the problem is formulated as
\begin{subequations}
\begin{align}
\mathcal{P}1: &~~~~\underset{{{\mathbf{S}},{\mathbf{V}},{{\left\| {{{\mathbf{w}}_R}} \right\|}_2^2} = 1}}{\text{max}}&& \!\!\!\!\!\!\!\!\!\!\!\!\!\!\!\!\!\!\!\!\!\!\!\!\!\!\!R_D^{\sec }({\mathbf{S}},{\mathbf{V}}) + {R_U^{sec}}({\mathbf{S}},{\mathbf{V}},{{\mathbf{w}}_R}) \\
&~~~~~~~~\text{s. t.}&&\!\!\!\!\!\!\!\!\!\!\!\!\!\!\!\!\!\!\!\!\!\!\!\!\!\!\!\operatorname{Tr} ({\mathbf{S}}) + \operatorname{Tr} ({\mathbf{V}}) \leqslant {P_{BS}},\\
&&& \!\!\!\!\!\!\!\!\!\!\!\!\!\!\!\!\!\!\!\!\!\!\!\!\!\!\!\!\!\!\!\!\!\!\!\!\!\!\!
\zeta \left( {{\mathbf{h}}_I^H{\mathbf{S}}{{\mathbf{h}}_I} + {\mathbf{h}}_I^H{\mathbf{V}}{{\mathbf{h}}_I} + {P_U}{{\left| {{g_I}} \right|}^2}} \right) \geqslant {E_{min}},
\end{align}
\end{subequations}
where $P_{BS}$ is the maximum power at the FD-BS and $E_{min}$ is the minimum requirement of the harvested energy at  $U_I$.

Notice that the feasible condition of problem $\mathcal{P}$1 is that ${E_{min}} \leqslant \zeta \left( {{P_{BS}}\left\| {{{\mathbf{h}}_I}} \right\|_2^2 + {P_U}{{\left| {{g_I}} \right|}^2}} \right)$ \cite{zhang2013mimo}. Throughout this letter, we consider the non-trivial case where  the  optimization problem is feasible. Obviously, problem $\mathcal{P}$1 is a non-convex problem. Therefore, the key idea to solve problem $\mathcal{P}$1 is the reformulation of the objective function.


\section{Optimization for The Sum  of Downlink and Uplink Secrecy Rate}

Observe that constraints (12b) and (12c) depend on variables $\mathbf{S}$ and $\mathbf{V}$, while constraint ${{\left\| {{{\mathbf{w}}_{R}}} \right\|}_2^2} = 1$ depends on variable $\mathbf{w}_R$. That is, constraints (12b) and (12c) are independent with ${{\left\| {{{\mathbf{w}}_{R}}} \right\|}_2^2} = 1$. So we can solve problem $\mathcal{P}$1 by first maximizing over ${{\mathbf{w}}_R}$, and then maximizing over $\mathbf{S}$ and $\mathbf{V}$ \cite{boyd2004convex}.

 Given the fixed $\mathbf{S}$ and $\mathbf{V}$, the optimization of problem $\mathcal{P}1$ is equivalent to  maximize the  uplink rate by finding the optimal receiver vector ${{\mathbf{w}}_R}$. Considering that the uplink channel is a SIMO channel,  the optimal unit receiver vector to maximize the SINR $\gamma_U$ is expressed as \cite{tse2005fundamentals}
\begin{equation}
{{\mathbf{w}}_R} = \frac{{{{\left( {\sigma _Z^2{{\mathbf{I}}_{{N_R}}} + {{\mathbf{H}}_{SI}}({\mathbf{S}} + {\mathbf{V}}){\mathbf{H}}_{SI}^H} \right)}^{ - 1}}{{\mathbf{g}}_U}}}
{{\left\| {{{\left( {\sigma _Z^2{{\mathbf{I}}_{{N_R}}} + {{\mathbf{H}}_{SI}}({\mathbf{S}} + {\mathbf{V}}){\mathbf{H}}_{SI}^H} \right)}^{ - 1}}{{\mathbf{g}}_U}} \right\|}_2}.
\end{equation}
Then, the  uplink  SINR at FD-BS is rewritten as
\begin{equation}
\begin{split}
 {\gamma_U}={P_U}{\mathbf{g}}_U^H{{\left( {\sigma _Z^2{{\mathbf{I}}_{{N_R}}} + {{\mathbf{H}}_{SI}}({\mathbf{S}} + {\mathbf{V}}){\mathbf{H}}_{SI}^H} \right)}^{ - 1}}{{\mathbf{g}}_U}.
\end{split}
\end{equation}
Substituting (14) into (10), ${R_U^{sec}}({\mathbf{S}},{\mathbf{V}},{{\mathbf{w}}_R})$ is consequently changed as ${R_U^{sec}}({\mathbf{S}},{\mathbf{V}})$. Then, the  problem $\mathcal{P}$1 can be further formulated as follows.
\begin{subequations}
\begin{align}
\mathcal{P}2:~~~~~ &\underset{{{\mathbf{S}},{\mathbf{V}}}}{\text{max}}&& \!\!\!\!\!\!\!\!\!\!\!\!\!\!\!\!\!\!\!\! \!\!\!\!\!\!\!\!\!\!R_D^{\sec }({\mathbf{S}},{\mathbf{V}}) + {R_U^{sec}}({\mathbf{S}},{\mathbf{V}}) \\
&\text{s. t.}&&\!\!\!\!\!\!\!\!\!\!\!\!\!\!\!\!\!\!\!\!\!\!\!\!\!\operatorname{Tr} ({\mathbf{S}}) + \operatorname{Tr} ({\mathbf{V}}) \leqslant {P_{BS}},\\
&&& \!\!\!\!\!\!\!\!\!\!\!\!\!\!\!\!\!\!\!\!\!\!\!\!\!\!\!\!\!\!\!\!\!\!\!
\zeta \left( {{\mathbf{h}}_I^H{\mathbf{S}}{{\mathbf{h}}_I} + {\mathbf{h}}_I^H{\mathbf{V}}{{\mathbf{h}}_I} + {P_U}{{\left| {{g_I}} \right|}^2}} \right) \geqslant {E_{min}}.
\end{align}
\end{subequations}

Although we have fixed the receiver vector, the objective function of problem $\mathcal{P}$2 is still complicated and non-convex. It is of importance to transform this problem into a tractable form. Motivated by the log-exponential reformulation idea in \cite{li2013coordinated}, we introduce slack variables $\{ {x_D},{y_D},{x_I},{y_I},{t_U},{y_U}\}$  to rewrite the problem $\mathcal{P}$2 as the following $\mathcal{P}$2.1:
\begin{subequations}
\begin{align}
\!\!\!\!\!\!&\underset{{{\mathbf{S}},{\mathbf{V}},{x_D},{y_D},\atop {x_I},{y_I},{t_U},{y_U}}}{\text{max}}&&\!\!\!\!\!\!\!\!\!\! \!\!\!\!\!\!\!\!\!({x_D} - {y_D} - {x_I} + {y_I} + {t_U}-{x_I}+{y_U}){\log _2}e \\
&~~~\text{s. t.}&&\!\!\!\!\!\!\!\!\!\!\!\!{\mathbf{h}}_D^H{\mathbf{S}}{{\mathbf{h}}_D} + {\mathbf{h}}_D^H{\mathbf{V}}{{\mathbf{h}}_D} +{P_U}{{\left| {{g_D}} \right|}^2}+ \sigma _Z^2 \geqslant {e^{{x_D}}},\\
&&&\!\!\!\!\!\!\!\!\!\!\!\!{\mathbf{h}}_D^H{\mathbf{V}}{{\mathbf{h}}_D} +{P_U}{{\left| {{g_D}} \right|}^2}+ \sigma _Z^2 \leqslant {e^{{y_D}}},\\
&&&\!\!\!\!\!\!\!\!\!\!\!\!{\mathbf{h}}_I^H{\mathbf{S}}{{\mathbf{h}}_I} + {\mathbf{h}}_I^H{\mathbf{V}}{{\mathbf{h}}_I} + {P_U}{\left| {{g_I}} \right|^2} + \sigma _Z^2 \leqslant {e^{{x_I}}},\\
&&&\!\!\!\!\!\!\!\!\!\!\!\!{\mathbf{h}}_I^H{\mathbf{V}}{{\mathbf{h}}_I} + {P_U}{\left| {{g_I}} \right|^2} + \sigma _Z^2 \geqslant {e^{{y_I}}},\\
&&&\!\!\!\!\!\!\!\!\!\!\!\!({x_D} - {y_D} - {x_I} + {y_I}){\log _2}e \geqslant 0, \\
&&&\!\!\!\!\!\!\!\!\!\!\!\!{P_U}{\mathbf{g}}_U^H{\left( {\sigma _Z^2{{\mathbf{I}}_{{N_R}}} + {{\mathbf{H}}_{SI}}({\mathbf{S}} + {\mathbf{V}}){\mathbf{H}}_{SI}^H} \right)^{ - 1}}{{\mathbf{g}}_U} \geqslant {e^{{t_U}}} - 1,\\
&&&\!\!\!\!\!\!\!\!\!\!\!\! {\mathbf{h}}_I^H{\mathbf{S}}{{\mathbf{h}}_I} +{\mathbf{h}}_I^H{\mathbf{V}}{{\mathbf{h}}_I}+ \sigma _Z^2 \geqslant {e^{{y_U}}},\\
&&&\!\!\!\!\!\!\!\!\!\!\!\! (t_U-x_I+y_U){\log _2}e \geqslant 0, \\
&&&\!\!\!\!\!\!\!\!\!\!\!\!\operatorname{Tr} ({\mathbf{S}}) + \operatorname{Tr} ({\mathbf{V}}) \leqslant {P_{BS}},\\
&&&\!\!\!\!\!\!\!\!\!\!\!\!\zeta \left( {{\mathbf{h}}_I^H{\mathbf{S}}{{\mathbf{h}}_I} + {\mathbf{h}}_I^H{\mathbf{V}}{{\mathbf{h}}_I} + {P_U}{{\left| {{g_I}} \right|}^2}} \right) \geqslant {E_{min}}.
\end{align}
\end{subequations}

 The objective function  (15a) is equivalently decomposed into the objective function  (16a) and the eight constraints of (16b)-(16i). In particular,  except for (16f) and (16i), the remaining six constraints will hold with equality at the optimum. If (16b) is not active at the optimality point, one can increase $x_D$ with a very small value, which improves the objective value while keeping other constraints unchanged. This contradicts the optimality point assumption. Other constraints can be proved in the same way. In addition, the  constraints (16f) and (16i) are to guarantee the non-negative secrecy rates of downlink and uplink, respectively.
 Hence, problem $\mathcal{P}$2.1 is equivalent to $\mathcal{P}$2.

However, the problem $\mathcal{P}$2.1 is still non-convex due to (16c), (16d) and (16g). In order to solve it efficiently,  we resort to an iterative algorithm based on sequential parametric convex approximation (SPCA) method \cite{beck2010sequential} to find an approximate solution. The non-convex parts of these constraints are iteratively linearized to its first-order Tayor expansion.

 To show this, let us first tackle the non-convex constraints (16c) and (16d). Suppose that, at iteration $n$, ${\mathbf{S}^*[n-1]}$, ${\mathbf{V}^*[n-1]}$, ${y_D^*}[n-1]$ and ${ x_I^*}[n-1]$ are given. A concave lower bound of $e^{y_D}$  in (16c) can be found as its first order approximation at a neighborhood of ${y_D^*}[n-1]$ because of  the  convexity of $e^{y_D}$. That is to say,
\begin{equation}
{e^{{{ y}_D^*}[n-1]}}({y_D} - { y_D^*}[n-1] + 1) \leqslant {e^{{y_D}}}
\end{equation}
 holds, implying that the approximation is conservative for (16c). Similarly,  we can replace ${e^{{x_I}}}$ by its conservative first order approximation ${{e^{{{ x}_I^*}[n-1]}}({x_I} - { x_I^*}[n-1] + 1)}$ in (16d).

  Then, we turn our attention to (16g). From \cite{nguyen2014spectral}, we know that ${P_U}{\mathbf{g}}_U^H{\left( {\sigma _Z^2{{\mathbf{I}}_{{N_R}}} + {{\mathbf{H}}_{SI}}({\mathbf{S}} + {\mathbf{V}}){\mathbf{H}}_{SI}^H} \right)^{ - 1}}{{\mathbf{g}}_U}$ is also joint convex with respect to ${\mathbf{S}}$ and ${\mathbf{V}}$, which is proved by epigraph and Schur complement. For ease of description, let  ${{\mathbf{X}}_U} = \sigma _Z^2{{\mathbf{I}}_{{N_R}}} + {{\mathbf{H}}_{SI}}({\mathbf{S}} + {\mathbf{V}}){\mathbf{H}}_{SI}^H$ and ${{{\mathbf{ X^*_U}}}}[n-1] = \sigma _Z^2{{\mathbf{I}}_{{N_R}}} + {{\mathbf{H}}_{SI}}({\mathbf{S}^*[n-1]} + {\mathbf{V}^*[n-1]}){\mathbf{H}}_{SI}^H$. Further define $G({{\mathbf{X}}_U},{{{\mathbf{X^*_U}}}}[n-1])$ as the first order approximation  of ${P_U}{\mathbf{g}}_U^H{{{\mathbf{X}}_U}^{ - 1}}{{\mathbf{g}}_U}$. In the same spirit as before, we have
  \begin{equation}
  \begin{split}
  &G({{\mathbf{X}}_U},{{{\mathbf{ X^*_U}}}}[n-1]) = {P_U}{\mathbf{g}}_U^H{{{\mathbf{ X^*_U}}}}{[n-1]^{ - 1}}{{\mathbf{g}}_U}\\
   &- \operatorname{Tr} \Big[ \left( {{P_U}{{{\mathbf{ X^*_U}}}}{{[n-1]}^{ - 1}}{{\mathbf{g}}_U}{\mathbf{g}}_U^H{{{\mathbf{ X^*_U}}}}{{[n-1]}^{ - 1}}} \right)\\
   &\left( {{{\mathbf{X}}_U} - {{{\mathbf{ X^*_U}}}}[n-1]} \right) \Big]
   \leqslant {P_U}{\mathbf{g}}_U^H{ {\mathbf{X}}_U^{ - 1}}{{\mathbf{g}}_U}.
  \end{split}
  \end{equation}
  To derive (18), we have used the fact that ${\nabla _{\mathbf{A}}}{{\mathbf{a}}^H}{{\mathbf{A}}^{ - 1}}{\mathbf{b}} =  - {{\mathbf{A}}^{ - 1}}{\mathbf{a}}{{\mathbf{b}}^H}{{\mathbf{A}}^{ - 1}}$ for ${\mathbf{A}} \succeq \mathbf{0}$ \cite{dattorro2008convex}.

Consequently, the convex approximate problem at iteration $n$ is the following problem $\mathcal{P}$2.2 :
\begin{subequations}
\begin{align}
\!\!\!\!\!\!&\underset{{{\mathbf{S}},{\mathbf{V}},{x_D},{y_D},\atop {x_I},{y_I},{t_U},{y_U}}}{\text{max}}&& \!\!\!\!\!\!\!\!\!({x_D} - {y_D} - {x_I} + {y_I} + {t_U}-{x_I}+{y_U}){\log _2}e \\
&~~~~\text{s. t.}&&\!\!\!\!\!\!\!\!{\mathbf{h}}_D^H{\mathbf{V}}{{\mathbf{h}}_D} +{P_U}{{\left| {{g_D}} \right|}^2} + \sigma _Z^2  \leqslant {e^{{{ y}^*_D}[n-1]}}({y_D} - {{ y}^*_D}[n-1] + 1),\\
&&&\!\!\!\!\!\!\!\!\!\!\!\!\!\!\!\!{\mathbf{h}}_I^H{\mathbf{S}}{{\mathbf{h}}_I} + {\mathbf{h}}_I^H{\mathbf{V}}{{\mathbf{h}}_I} + {P_U}{\left| {{g_I}} \right|^2} + \sigma _Z^2 \leqslant {e^{{{ x}^*_I}[n-1]}}({x_I} - {{ x}^*_I}[n-1] + 1),\\
&&&\!\!\!\!\!\!\!\!\!\!\!\!\!\!\!\!G({{\mathbf{X}}_U},{{{\mathbf{ X}}}^*_U}[n-1]) \geqslant {e^{{t_U}}} - 1, \displaybreak[0]\\
&&&\!\!\!\!\!\!\!\!\!\!\!\!\!\!\!\!\text{(16b), (16e), (16f), (16h)-(16k)}.\
\end{align}
\end{subequations}
This is a convex SDP which can be solved efficiently by off-the-shelf solvers, e.g., CVX \cite{Grant2012cvx}. By solving this problem, we can obtain ${\mathbf{S}^*[n]}$, ${\mathbf{V}^*[n]}$, ${y_D^*}[n]$, ${ x_I^*}[n]$ as well as the achieved sum secrecy rate $u[n]$. Detailed steps to solve problem $\mathcal{P}$2 are stated in Algorithm 1. According to \cite{beck2010sequential}, Algorithm 1  converges to a KKT point of problem $\mathcal{P}2$. Detailed proof is presented in Appendix A. It is worth noting that the iterative procedure in Algorithm 1 may return a locally optimal solution to problem $\mathcal{P}$2.

\begin{algorithm}[h]
\caption{  SPCA method for problem $\mathcal{P}$2}
\begin{algorithmic}[1]
    \STATE {Initialize feasible points for ${\mathbf{S}^*[0]}$ and ${\mathbf{V}^*[0]}$ by solving the feasibility problem of $\mathcal{P}$2 (replace (15a) with 0);
    \STATE Calculate  ${{ y}^*_D}[0] = \ln ({\mathbf{h}}_D^H{\mathbf{V}^*[0]}{{\mathbf{h}}_D} +{P_U}{{\left| {{g_D}} \right|}^2}+ \sigma _Z^2)$} and ${{ x}^*_I}[0] = \ln ({\mathbf{h}}_I^H{\mathbf{S}^*[0]}{{\mathbf{h}}_I} + {\mathbf{h}}_I^H{\mathbf{V}^*[0]}{{\mathbf{h}}_I} + {P_U}{\left| {{g_I}} \right|^2} + \sigma _Z^2)$;
    \STATE Set $n:=0$;
    \WHILE {$\frac{{u[n] - u[n - 1]}}{{u[n - 1]}} \geqslant {10^{ - 3}}$}
    \STATE Solve problem $\mathcal{P}$2.2 by CVX to obtain ${\mathbf{S}^*[n]}$, ${\mathbf{V}^*[n]}$, ${y_D^*}[n]$ and ${ x_I^*}[n]$;
    \STATE Set $n:=n+1$;
    \ENDWHILE
    \RETURN ${\mathbf{S}^*[n]}$ and ${\mathbf{V}^*[n]}$ as an approximate solution.
    \label{code:recentEnd}
 \end{algorithmic}
\end{algorithm}


\section{Simulation Results}

In this section, simulation results are presented to evaluate the performance of our proposed schemes. We assume that FD-BS is equipped with $N_T=4$ and $N_R=4$ antennas and its transmission power  is ${P_{BS}} = 1$ W. The uplink user transmission power is  0.1 W. For simplicity, we set the energy harvesting efficiency as 50$\%$. All the receiver noise power equals to $-80$ dB. Assume that the signal attenuation from FD-BS to idle user is 30 dB and the remaining channel attenuations are 70 dB excluding the residual SI channel. These channel entries are independently generated from i.i.d Rayleigh fading with the respective average power values. Besides, we generate the elements of ${{\mathbf{H}}_{SI}}$ as ${\cal CN}(0,\sigma _{SI}^2)$, where $\sigma _{SI}^2$ depends on the capability of the SIC techniques.
For comparison, we also introduce two  schemes, i.e., perfect full duplex and two-phase half duplex. In the half duplex scheme, all $N=8$ antennas are used for data transmission/reception in 1/2 time slot.


\begin{figure}
\centering
\includegraphics[width=9cm, height=7.8cm]{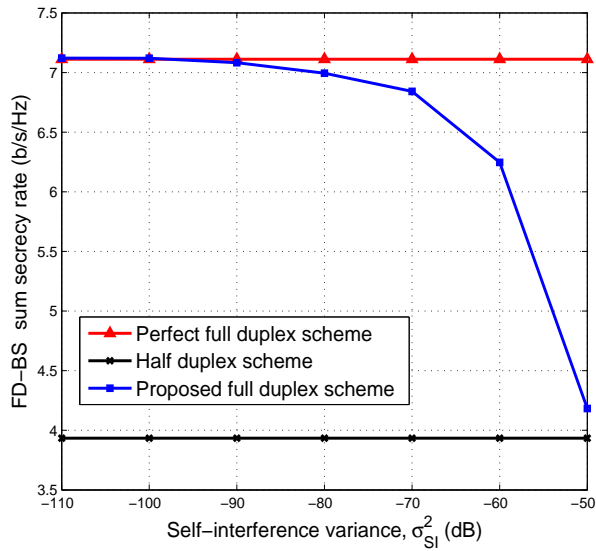}
\caption{Achievable sum secrecy rate versus self-interference variance for different schemes with  $E_{min}=1$ mW.}
\label{systemmodel}
\end{figure}

In Fig. 2, the impact of self-interference variance on the achievable sum secrecy rate is presented with $E_{min}=1$ mW. As expected, we can see that the performance of our proposed full duplex scheme  degrades as $\sigma_{SI}^2$ increases, while that of other schemes  remain the same.

\begin{figure}
\centering
\includegraphics[width=9cm, height=7.8cm]{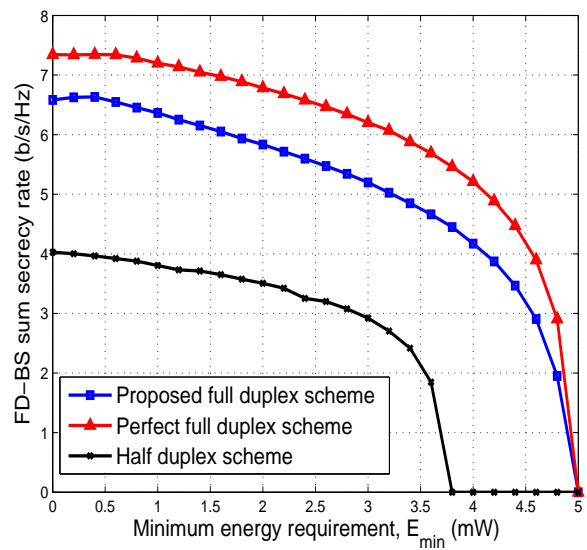}
\caption{Achievable sum secrecy rate versus minimum energy requirement for different schemes with $\sigma_{SI}^2=-60$ dB.}
\label{systemmodel}
\end{figure}

Fig. 3 compares the sum secrecy rate of different schemes versus minimum energy requirement  with $\sigma_{SI}^2=-60$ dB. According to  Fig. 3, it is straightforward that the secrecy sum rate decreases as the minimum energy requirement increases. Besides, the performance of our proposed full duplex scheme extremely outperforms that of half duplex scheme by 65\%,  which greatly verifies the superiority of our proposed system.

%

\section{Conclusion}

This letter has designed a transceiver scheme for  secure SWIPT in a full duplex wireless system with one downlink user, one uplink user and one idle user.    We have  jointly optimized the transmitter covariance matrix and receiver vector to maximize the sum of downlink and uplink secrecy rate subject to the sum transmission power constraint at FD-BS and the harvested energy requirement at the idle user. It has been demonstrated that  the proposed full duplex scheme yields a large gain than   half duplex scheme in terms of the sum secrecy rate.



\appendices
\section{Proof of the convergence for Algorithm 1}
In this appendix, we first prove the convergence of Algorithm 1 by adopting the technique from  \cite{beck2010sequential}.  Let ${\mathcal{S}}^{(n)}$ be the convex set of problem $\mathcal{P}$2.2 at iteration $n$. For ease of presentation,  we define
\begin{equation}
\varphi ({y_D},{\mathbf{V}}) = {\mathbf{h}}_D^H{\mathbf{V}}{{\mathbf{h}}_D}+{P_U}{{\left| {{g_D}} \right|}^2} + \sigma _Z^2 - {e^{{y_D}}} ~~\text{and}
 \end{equation}
  \begin{equation}
  \begin{split}
  &\phi ({y_D},y_D^*[n - 1],{\mathbf{V}}) = {\mathbf{h}}_D^H{\mathbf{V}}{{\mathbf{h}}_D} +{P_U}{{\left| {{g_D}} \right|}^2}+ \sigma _Z^2 \\
  &~~~~~~~- {e^{y_D^*[n - 1]}}({y_D} - y_D^*[n - 1] + 1).
   \end{split}
   \end{equation}
   Due to (17) and (19b), we have
   \begin{equation}
   \varphi({y_D},{\mathbf{V}})  \leqslant \phi ({y_D},y_D^*[n - 1],{\mathbf{V}}) \leqslant 0.
   \end{equation}
Since the affine approximation in (17), the following  two properties are satisfied.
\begin{align}
&{ {\phi ({y_D},y_D^*[n - 1],{\mathbf{V}})} \bigg|_{{y_D} = y_D^*[n - 1],{\mathbf{V}} = {{\mathbf{V}}^*}[n - 1]}} \nonumber \\
& \!\!\!\!\!\!\!=  { {\varphi ({y_D},{\mathbf{V}})} \bigg|_{{y_D} = y_D^*[n - 1],{\mathbf{V}} = {{\mathbf{V}}^*}[n - 1]}} \leqslant 0,
\end{align}
\begin{align}
&{ {\nabla \phi ({y_D},y_D^*[n - 1],{\mathbf{V}})} \bigg|_{{y_D} = y_D^*[n - 1],{\mathbf{V}} = {{\mathbf{V}}^*}[n - 1]}} \nonumber \\
&\!\!\!\!\!\!\!\!= { {\nabla \varphi ({y_D},{\mathbf{V}})} \bigg|_{{y_D} = y_D^*[n - 1],{\mathbf{V}} = {{\mathbf{V}}^*}[n - 1]}}.
\end{align}
From (23), we know that the optimal variables $y_D^*[n - 1]$ and $\mathbf{V}^*[n - 1]$ obtained at iteration $n-1$ is a feasible solution to the problem $\mathcal{P}$2.2 at iteration $n$.
Similarly,  the constraints in (19c) and (19d) also have the same properties. In other words,
$  \left( y_D^*[n - 1], x_I^*[n - 1], \mathbf{S}^*[n - 1], \mathbf{V}^*[n - 1]\right) \in \mathcal{S}^{(n)}$ and thus $u[n] \geqslant u[n-1]$. In fact, we have shown that the sequence ${u[n]}$ in nondecreasing. Besides, the value of ${u[n]}$ is bounded above due to the limited transmission power, and thus it is guaranteed to convergence.

Then, similar with that (19b) has two properties, i.e., (23) and (24), (19c) and (19d) also have their corresponding properties, respectively.
According to [15, Proposition 3.2],  all accumulation points of  $ \left( y_D^*[n], x_I^*[n], \mathbf{S}^*[n], \mathbf{V}^*[n]\right)$ are KKT points of the original problem $\mathcal{P}$2.1 or $\mathcal{P}$2. Thus, our proposed algorithm 1 converges to a KKT point of problem $\mathcal{P}$2.

\ifCLASSOPTIONcaptionsoff
  \newpage
\fi



%

\bibliographystyle{IEEEtran}

\bibliography{reference}
%

%
%
%




\end{document}